\documentclass[a4paper]{jpconf}
\usepackage[cp1252]{inputenc}
\usepackage{iopams}
\usepackage{cite}
\usepackage{graphicx}
\begin{document}
\title{Dynamics of supported ultrathin molybdenum films driven by strong short laser impact}
\author{V~A~Khokhlov$^{1}$, Yu~V~Petrov$^{1,2}$, N~A~Inogamov$^{1,3}$, K~P~Migdal$^{3,1}$,
J~Winter$^{4,5,6}$, C~Aichele$^{4}$, S~Rapp$^{4,5,6}$ and H~P~Huber$^{4}$
}

\address{$^1$ Landau Institute for Theoretical Physics of the Russian Academy of Sciences, Akademika Semenova 1a, Chernogolovka, Moscow Region 142432, Russia}
\address{$^2$ Moscow Institute of Physics and Technology, Institutskiy Pereulok 9, Dolgoprudny, Moscow Region 141700, Russia}
\address{$^3$ Dukhov Research Institute of Automatics (VNIIA), Sushchevskaya 22, Moscow 127055, Russia}
\address{$^4$ Department of Applied Sciences and Mechatronics, Munich University of Applied Sciences,
Lothstrasse 34, Munich 80335, Germany}
\address{$^5$ Erlangen Graduate School in Advanced Optical Technologies ,
Friedrich-Alexander-Universitr{\"a}t Erlangen-Nr{\"u}rnberg, Paul-Gordan-Strasse 6, Erlangen 91052, Germany}
\address{$^6$ Lehrstuhl fr{\"u}r Photonische Technologien, Friedrich-Alexander-Universitr{\"a}t
Erlangen-Nr{\"u}rnberg, Konrad-Zuse-Strasse 3-5, Erlangen 91052, Germany}

\ead{nailinogamov@gmail.com}

\begin{abstract}
We consider expansion, break off, and flight of 10 nm molybdenum film deposited
onto glass support. These events are initiated by action of subpicosecond laser pulse onto film.
Approximations for two-temperature equation of state and electron--ion coupling parameter
are developed. Heat conduction is unimportant because film is ultrathin and because radius
of a laser beam is rather large $\sim 10$ $\mu$m (thus lateral thermal spreading is insignificant at
the considered time scale). We use two-temperature one-dimensional hydrodynamic code to
follow evolution of laser induced flow. Additional code for treating transmission and reflection
of a monochromatic electromagnetic wave is developed. It is applied to describe interference
between transmitted and reflected waves in the layered structure appearing thanks to laser
induced expansion and separation of a film.
\footnote{Accepted to the Journal of Physics: Conference Series}
\end{abstract}

\section{Introduction}\label{intro}

\section{Thermodynamic functions of molibdenum}
Important direction in physics of laser--matter interaction is connected with ultrashort laser
pulses (UsLP). There are three ranges of their fluences $F$: weak, moderate, and strong. The
range of moderate fluences begins around melting threshold $F_m$ for bulk metal targets and
continues up to $\sim (10^2$--$10^3) F_m$, see \cite{Inogamov:2009a}. 
In absorbed fluences $F_{abs}$, the melting threshold $F_{abs}|_{m}$
is of the order of few tens of mJ/cm$^2$ for bulk metal targets. There are two subranges of high
fluences above the moderate range. They are below \cite{Meyer-ter-Vehn:2000:PRE,Agranat:2007} 
and above \cite{AndreevKuznetsov,Esirkepov:2004:PRL,Bychenkov:2014:LPB} the relativistic intensity
$\sim 10^{18}$--$10^{19}$ W/cm$^2$. At this intensity the quivering kinetic energy of electron oscillations in
a laser electromagnetic wave $90\lambda_1^2 I_{18}$ keV becomes of the order of the rest energy of electron
while amplitude of electron oscillations in vacuum $1000\lambda_1^2 \sqrt{I_{18}}$ nm becomes larger than thickness
$d_{skin} = 10$--20 nm of a skin-layer, here $\lambda_1 = \lambda /1$ $\mu$m is wavelength of a wave, 
$I_{18} = I/10^{18}$ W/cm$^2$ is its intensity.

Below we consider the moderate range of laser actions. With UsLP new physics of laser
ablation appears. This is the thermomechanical ablation. It differs qualitatively from previously
known ablation by subnanosecond and nanosecond pulses which are of the purely evaporative
origin. The first manifestation (1998, 1999) about the principal change of the origin comes from
observation of Newton rings rising their number with time \cite{Sokolowski-Tinten:PRL:1998,InogamovJETPLett:1999,Zhakhovskii:2000}. Thermomechanical ablation was also observed in molecular dynamics (MD) simulations of laser heating of soft tissues \cite{Zhigilei:2000}.

Duration $\tau_L$ of UsLP is less than $\sim 1$ ps. In this case we have to include into consideration
a finite time relaxation between electron and ion subsystems of a metal \cite{SIA:1974}. During UsLP
the laser energy is absorbed by electrons. Thus electron subsystem become hotter than ion
one: $T_e > T_i$, where $T_e$ and $T_i$ are electron and ion temperatures. This is the two-temperature
(2T) case important for UsLP. Thanks to electron--ion coupling the temperatures $T_e$ and $T_i$
equalize gradually \cite{SIA:1974,Hohlfeld:2000} during equilibration time $t_{eq}$. It should be mentioned that duration
of equilibration teq depends on thickness of a target df relative to thickness $d_T$ $(d_T > d_{skin})$ of
a heat affected zone in a bulk target. For thick targets $d_f > d_T$ the electron temperature $T_e$ in
large extent drops down thanks to electron heat conduction from a skin $d_{skin}$ to a heat affected
zone $d_T$; $d_T > d_{skin}$. While for thin films $d_f \ll d_T$ the conductive loses are insignificant, and
duration $t_{eq}$ is longer, because only electron--ion coupling contributes to decrease of $T_e$ and to
drawing together the temperatures $T_e$ and $T_i$.

The first paper introduced the 2T states $T_e > T_i$ in laser irradiated condensed matter was
\cite{SIA:1974}. Significantly later on the laser systems based on chirped pulse amplification and generating
UsLPs appear \cite{Strickland:1985} and the approach with 2T states becomes popular \cite{Inogamov:2009a, Meyer-ter-Vehn:2000:PRE, Agranat:2007, Hohlfeld:2000, Allen-PRL-1987, Wang+Riffe+Downer-PRB-1994,Lugovskoy+Bray:1999:PRB, Rethfeld:2002:PRB, Petrov:2005, Zhibin:2008, JETP2010Inogamov} because
people began to use action of UsLP onto targets while the 2T approach allows them to describe
results of laser–matter interaction. Experiments in the nineties operated with rather weak
UsLPs, thus electrons were overheated above lattice to a few thousand Kelvins and electron
energy per unit of volume and pressure were rather small. Later on more strong actions became
possible, they cause overheating to a few tens of kiloKelvins and may cause the thermomechanical
ablation.

Inclusion of electronic contribution to thermodynamics of condensed matter was proposed in
the fifties--sixties \cite{PhysRev.75.1561, PhysRev.96.934, PhysRev.99.550, Altshuler:1960, Kormer:1662, Altshuler:1962}:
\begin{equation}\label{Eq:2T}
f = f_i + f_e,
\end{equation}
where $f$, $f_i$, and $f_e$ are free energies of whole system and its ionic and elecronic contributions.
This allows to develope the one-temperature (1T)
\begin{equation}\label{Eq:1T}
Te = Ti = T
\end{equation}
wide-range equations of states (EoS) for high temperatures where the electronic contribution
becomes significant thanks to excitation of electrons \cite{Fortov:2007:UFN, Fortov+Lomonosov:2014}. Let’s mention that the paper \cite{SIA:1974} breaks the condition (\ref{Eq:1T}) and separates independent temperatures. For low temperatures T in (\ref{Eq:1T})
the electronic excitations are weak and the electronic addition $f_e$ in (\ref{Eq:2T}) to EoS is insignificant.
What temperatures are low in the sense that $f \approx f_i \ll f_e$?

Critical temperatures $T_c$ of metals, especially rarefactory metals, are rather high. Let’s show
that at the temperatures $\approx T_c$ the contribution fe is important. Estimates based on Fermi
theory of free electrons show that
\begin{equation}\label{Eq:3}
P_e = \gamma T_e^2/3, \qquad c_e = \gamma T_e, \qquad \gamma  = \gamma_0 (n_e/n_{e0})^{1/3},
\end{equation}
here $c_e$ is a heat capacity of an electron subsystem per unit volume, $n_e = Zn$ is electron
concentration, $n$ is atomic concentration, $n_{e0}$ is electron concentration at solid state density. We
have
\begin{equation}\label{Eq:4}
n_c \approx  (1/3\hbox{--}1/4)n_0, \qquad T_c \approx 10^4 K, \qquad \gamma_0 \sim 100 J/m^3/ K^2.
\end{equation}
Using (\ref{Eq:3}) and (\ref{Eq:4}), supposing that the charge $Z$ remains fixed during this range of expansions,
and taking $Z = 1$, we obtain the electron pressure at critical parameters $P_e|_c \approx 2$ GPa.

While critical temperature of Mo is 11.8 kK according to [29]. It is slightly higher than
the isotherm 104 K. Quantum molecular dynamics (QMD) simulations \cite{1742-6596-946-1-012093} give unusually large
degree of expansion $\approx 6.5$ between critical and normal density points. Approximation (\ref{Eq:2T})
together with Mie--Gr{\"u}neisen approach implies that EoS may be split up into three contributions
\begin{equation}
f(\rho, T_e, T_i) = f_s(\rho) + \phi_i(\rho, T_i) + \phi_e(\rho, T_e).
\end{equation}
Thus separation of temperatures $T_i$ and $T_e$ essentially splits the Mie--Gr{\"u}neisen thermal
contribution $\phi$ into electron and ion parts together with the cold (static) contribution (we
neglect zero-point vibrational functions in comparison with the static functions and thermal
exitations). The contributions $\phi_i(\rho,  T_i)$ and $\phi_e(\rho, T_e)$ may be approximately factorized into
density and temperature dependent factors. Thus electron and ion Gr{\"u}neisen $\Gamma_e$
and $\Gamma_i$ appear.

Electron addition to pressure $P_e$ has a thermal nature. It much weaklier depends on density
relative to cold pressure. Therefore pressure $P_e$ remains significant at rather large (few times)
volume extensions, see (\ref{Eq:3}), (\ref{Eq:4}). Estimates (\ref{Eq:3}), (\ref{Eq:4}) give $P_e|_{\rm crit} \approx 2$ GPa. 
This estimate are few
times larger than critical pressure $P_c \approx 0.7$ GPa \cite{1742-6596-946-1-012093}. 
Negative ion pressures $P_i(\rho, T_i) < 0$ are
necessary to decrease total pressure to $P_c$.

Values $P_i(\rho, T_i) < 0$ are usual for metastable stretched states \cite{Kanel:2007} and for physics of UsLP
with 2T states. Let’s compare two phase coexistence boundaries (binodal, bin) for 1T (2) and
2T states:
\begin{eqnarray}
(\rho_{\rm bin-1T}, T_e = T_{\rm bin-1T}, T_i = T_{\rm bin-1T}), &&
(\rho_{\rm bin-2T}, T_{e\, {\rm bin-2T}}, T_{i\, {\rm bin-2T}} = T_{\rm bin-1T}),\label{Eq:6} \\
T_{e\, {\rm bin-2T}} > T_{\rm bin-1T}.\nonumber
\end{eqnarray}
Additional excitation of electrons rises their positive pressure which additionally expands ion
subsystem in equilibrium conditions where total pressure is small. Therefore the 2T coexistence
boundary (\ref{Eq:6}) is shifted to smaller densities
$$
\rho_{\rm bin-2T} < \rho_{\rm bin-1T}
$$
relative to the 1T coexistence boundary at the same ion temperature $T_i = T_{\rm bin-1T}$ and elevated
electron temperatures $T_{e\,{\rm bin-2T}} > T_{\rm bin-1T}$ \cite{Inogamov:2012a, Migdal:2015:Elbrus, Petrov:2015:APB, Ashitkov:2016:JPCC, Tanaka:2018}, see also figures 18--22 in \cite{Ilnitsky:2016}. Equilibrium
2T states which can contact with saturated vapor disappear for electron temperatures higher
than 20--25 kK at the any (even low $T_i = 0$) ion temperatures for all studied metals from Ni
and Ta to Au and Cu \cite{Inogamov:2012a, Migdal:2015:Elbrus, Petrov:2015:APB, Ashitkov:2016:JPCC, Ilnitsky:2016}. 
This corresponds to the near critical region of 2T EoS. Metal
transfers to the supercritical 2T fluid state above (that is at higher temperatures) this region.
Difference between liquid and vapor eliminates in the supercritical 2T states. It is supposed
that atoms transit from condensed to vapor states keeping their degree of electron excitation
during 2T evaporation. Problem of 2T saturated vapor needs special discussion.

In this paper Mie--Gr{\"u}neisen two-temperature approach is applied to describe twotemperature thermodynamical characteristics of molybdenum in the process of its ablation under
the action of femtosecond laser pulses. Molybdenum is the essential constructive element of the
pump--probe laser optical schemes. It refers to metals with a high melting point. In this
work we consider a thin molybdenum film (10 nm thickness) deposited on a glass substrate
(of 500 $\mu$m thickness). This design is a target for the femtosecond laser impulse within the
pump--probe scheme. Infrared probe pulse has a light wave length 1053 nm while the wave
length of the probe pulse is 527 nm. We are interesting in a interference pattern produced
by the probe pulse during the process of ablation after the action of the pump pulse. If we
consider this interference in the small vicinity of the center of laser spot on the target, the
ablation pattern with the sufficient accuracy can be represented as one-dimensional. 
This onedimensional ablative motion of a target matter we investigate with the use of a hydrodynamicalcode. This code calculates parameters of the moving matter by the use of its two-temperature
thermodynamics and two-temperature kinetics. State of the ablative matter, distribution of its
velocity, density, temperature defines the interference picture produced by the probe pulse. We
calculate the temporal evolution of this interference picture on the multilayer ablative formation.

\section{Thermodynamic functions of molybdenum}
We shall consider thermodynamic functions of molybdenum in the framework ofMie--Gr{\"u}neisen approach.
In the two-temperature state with the temperatures of ions $T_ i$ and the temperature of electrons $T_e$ internal energy will be presented as a sum
$$
E(x,T_i,T_e)=E_i (x, T_i)+E_e (x, T_ e),
$$
where its ionic part
$$
E_ i (x, T_i)=E_s (x) + E_T (x, T_i)
$$
consists of the static energy $E_ s(x)$ and the thermal part $E_T (x, T_i)$
Here we have inroduced the relative density of the material $x=\rho/\rho_0$, where $\rho_0$ is the density at zero temperature and pressure.
Acording to our calculations based on the density functional theory it corresponds to the volume per atom $v_0=105.47 a_{\rm B}^3$ ($a_{\rm B}$ is the Bohr radius).
Electronic part $E_e(x,T_e)$ describes the contribution of the excited electrons into the interal energy.

The static energy is chosen in the form consisting of the repulsive part at the small values of $x$ and interaction at large $x$ and as a quantity per atom is
$$
E_s(x)=A\left(\frac{x^a}{a}-\frac{x^b}{b}\right).
$$
with $a>b$.
And the thermal part of the ionic energy (per atom)is simply
$$
E_T (x, T_i)=3 k_{\rm B} T_i
$$
with $k_{\rm B}$ being the Boltzmann constant and the ion as the electron temperature to be measured in K.

Another thermodynamic function important at the hydrodynamics
research of the laser ablation of materials is the prassure which within the same approach is
$$
P(x,T_i,T_e)=P_i (x, T_ i)+P_e (x, T_e)
$$
with the ionic part
$$
P_(x, T_i)=P_s (x) + P_T (x, T_i).
$$
Here
\begin{equation}\label{Eq:Ps}
P_s (x)=\frac{A}{v_0}x(x^a-x^b)
\end{equation}
is the static pressure, and
$$
P_T (x, T_i)=\frac{3 k_{\rm B} T_i}{v_0}xG(x)
$$
is the thermal part of pressure.
Gr{\"u}neisen parameter $G(x)$ of the solid phase of metal is $G(x)=\rmd\ln \theta/\rmd \ln x$ with $\theta (x)$ being the Debye temperature. The Debye temperature $\theta=\hbar s(x)k_{\rm D}(x)/k_{\rm B}$, where $s(x)$ and $k_{\rm D} (x)$ are the sound speed and Debye wave number. Taking into account that $k_{\rm D}\propto x^{1/3}$ and
$$
s(x)\propto \sqrt{\frac{\partial P_{\rm s}}{\partial \rho}} \propto
\sqrt{\frac{\partial P_{\rm s}}{\partial x}} \propto
\sqrt{(a+1)x^a-(b+1)x^b},
$$
we obtain
\begin{equation}\label{Eq:Gr}
G(x)=\rmd\ln \theta/\rmd \ln x=\frac{1}{3}+\frac{1}{2}
\frac{a(a+1)x^a-b(b+1)x^b}{(a+1)x^a-(b+1)x^b}
\end{equation}
To avoid divergence of this expression at small x we change it to a close expression \cite{Petrov:2015:Elbrus}
\begin{equation}\label{Eq:Gr:9}
G(x)=\frac{5}{6} + \frac{2a(b+1) + (a-b)(a-1)x^{a+1}}{2(b+1) + (a - b)x^{a+1}}
\end{equation}
Functions (\ref{Eq:Gr}) and (\ref{Eq:Gr:9}) have equal values and equal values of their derivatives at x = 1.

The electron contribution to the internal energy and pressure $E_e (x, T_e)$ and $P_e (x, T_e)$ was calculated in the VASP program of the density functional method as the difference between the values of energy and pressure at electron temperature $T_e$ and at a low electron temperature. The results of these calculations for the electron internal energy per unit volume can be approximated in the form of expression
$$
E_e(x,T_e)=e_{e1} x^{5/3}(\sqrt{\tau^2+\tau_e^2}-\tau_e.
$$
Here we have introduced dimensionless variable $\tau = T_e/(T_{a6}x^{2/3})$ (electron temperature $T_e$ is
measured in K), $T_{a6} = 6T_{a1}$, $T_{a1} = 11605$ K, $e_{e1} = 244.9$ GPa, $\tau_e = 0.1508$. Accordingly
elecronic part of pressure can be approximated as
$$
P_e(x,T_e)=p_{e1}x^{5/3}(\sqrt{\tau^2+\tau_p^2}-\tau_p
$$
with $p_{e1} = 235.1$ GPa, $\tau_p = 0.08507$.

From the expression (\ref{Eq:Ps}) for the static pressure the minimal static pressure $-p_m$ can be found.
Minimumal value of the pressure (\ref{Eq:Ps}) takes place at the compression
$$
x_m = \left(\frac{b+1}{a+1}\right)^{\frac{1}{a-b}}
$$
and is equal to
$$
-p_m = \frac{A}{v_0} (b-a) \left[\frac{(b+1)^{b+1}}{(a+1)^{a+1}}\right]^{\frac{1}{a-b}}.
$$
Then the parameter A can be expressed through the $p_m$ value:
\begin{equation}\label{Eq:A:10}
A =  \frac{p_m v_0}{a-b} \left[\frac{(a+1)^{a+1}}{(b+1)^{b+1}}\right]^{\frac{1}{a-b}}.
\end{equation}
Our DFT calculations consistent with other data \cite{Duffy:1999, MoEOS} give $p_m = 40$ GPa. With the help of
(\ref{Eq:A:10}) there are only two free parameters---$a$ and $b$ for the ionic part of energy and pressure. They
were calculated to reproduce the zero pressure isobar obtained by the use of quantum molecular
dynamics in \cite{1742-6596-946-1-012093}. Considering the zero pressure isobar, we have $P = P_i + P_e = 0$. Substituting
here expressions for the ionic and electronic parts of pressure, in the equilibrium state with the
temperature $T = T_i = T_e$ we obtain
\begin{equation}\label{Eq:11}
\frac{A}{v_0} x(x^a - x^b) + \frac{3k_{\rm B}T}{v_0} xG(x) + 
p_1 x^{5/3} 
\left( \sqrt{ \left( \frac{k_{\rm B}T}{T_{a6}x^{2/3}}\right)^2 + \tau_p^2} - \tau_p \right) = 0
\end{equation}
By introducing the notations
\begin{eqnarray*}
\omega_1&=&\frac{A}{v_0}x(x^a - x^b),\\
\omega_2&=&\frac{3}{v_0},\\
\omega_3&=&p_1 x^{5/3},\\
\omega_4&=&\left(\frac{1}{T_{a6} x^{2/3}}\right)^2,
\end{eqnarray*}
we obtain from the equation (\ref{Eq:11}) for the dependence of the temperature $T$ on the compression
$x$ for the zero pressure isobar
\begin{equation}\label{Eq:KBT:12}
k_{\rm B}T = \frac{ \omega_2 (\omega_3 b_1 - \omega_1) -
                     [\omega_2^2 (\omega_3 b_1)^2 + \omega_1 (\omega_2^2 - \omega_3^2 \omega_4) (2\omega_3 b_1 - \omega_1)]^{1/2} }
                     {\omega_2^2 - \omega_3^2 \omega_4}.
\end{equation}
\begin{figure}
\centering 
\includegraphics[width=0.6\columnwidth]{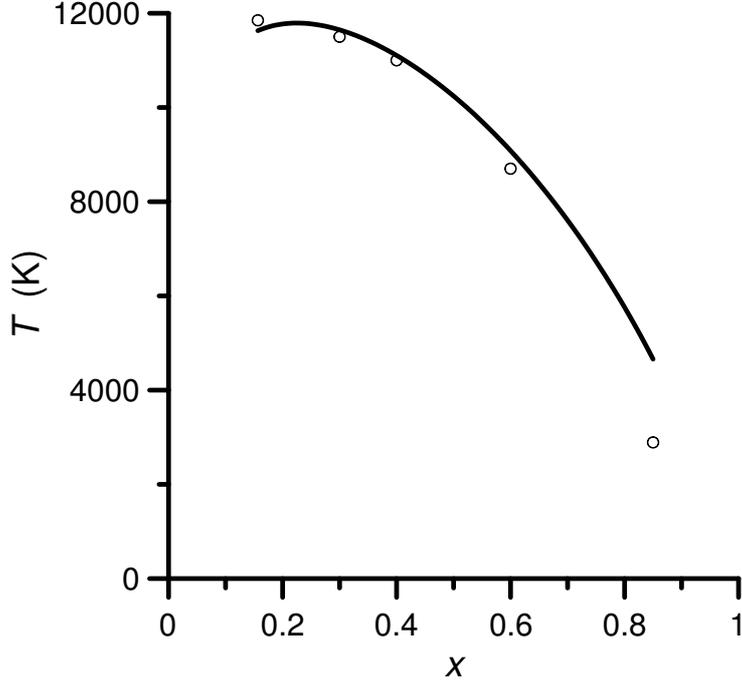}
\caption{Zero isobar of molybdenum obtained by the use of the equation (\ref{Eq:KBT:12}) (solid line) in
comparison with the zero isobar from the work \cite{1742-6596-946-1-012093} (circles).}\label{Fig:Pzero}
\end{figure}
\begin{figure}
\centering 
\includegraphics[width=0.6\columnwidth]{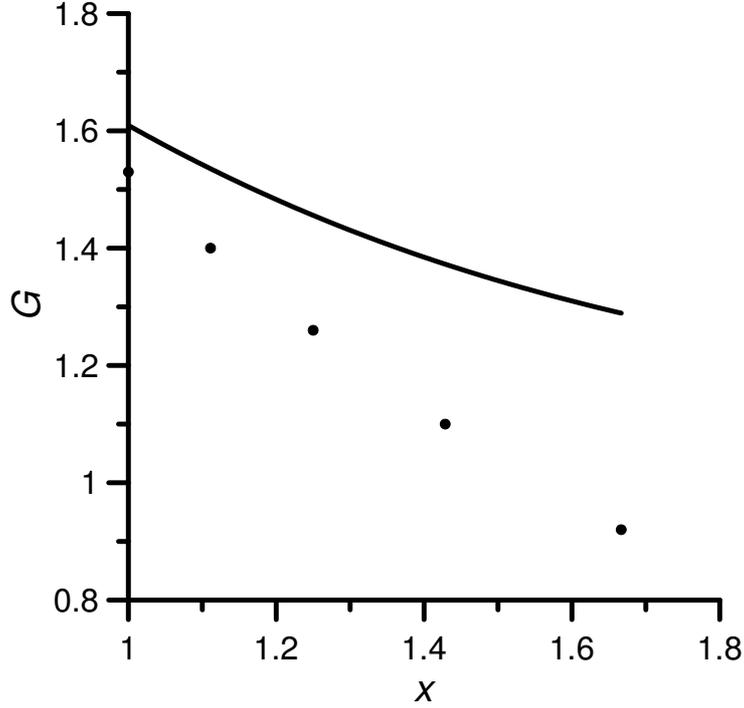}
\caption{Gr{\"u}neisen parameter of molybdenum (\ref{Eq:Gr:9}) in dependence on the compression x. Circles
are the results obtained in \cite{Fatyanov:2017:JAP}.}\label{Fig:Grun}
\end{figure}
From the comparison with the zero pressure isobar obtained in \cite{1742-6596-946-1-012093} we find values a and b:
$$
a = 1.376, \qquad b = 0.176.
$$
Then $A = 0.556$ eV from (\ref{Eq:A:10}). 
Zero pressure isobar as the dependence of the temperature $T$
on the compression $x$ is presented in figure \ref{Fig:Pzero}. In figure \ref{Fig:Grun} theGr{\"u}neisen parameter (\ref{Eq:Gr:9}) with the
parameters $a$ and $b$ found is shown together with that one calculated in \cite{Fatyanov:2017:JAP}. 
In figure \ref{Fig:Tcrit} several
isotherms close to the critical isotherm are drawn. Critical parameters corresponds to
\begin{figure}
\centering 
\includegraphics[width=0.6\columnwidth]{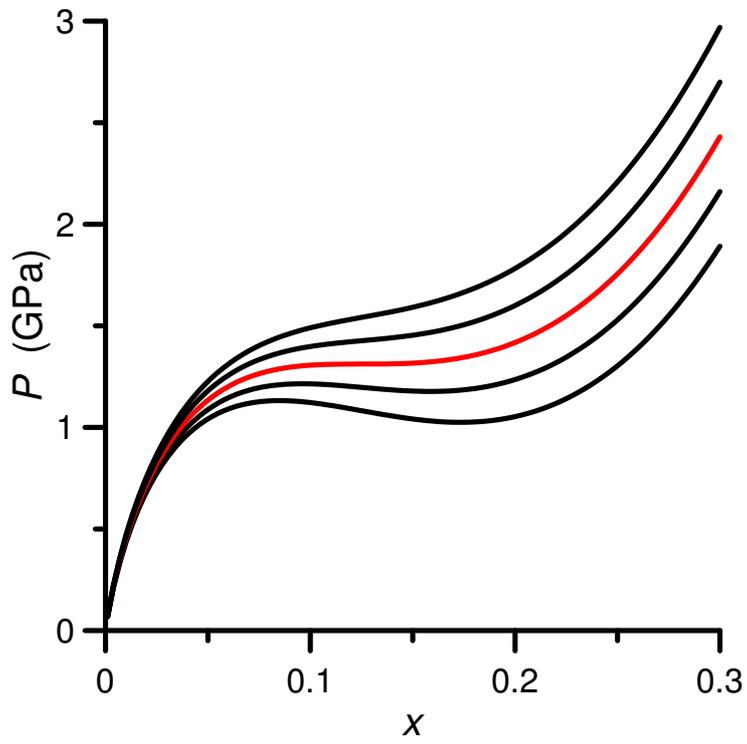}
\caption{Isotherms of molybdenum close to its critical isotherm (red line). Temperatures at
isotherms down up are T=123500, 12450, 12550, 12650, 12750 K.}\label{Fig:Tcrit}
\end{figure}
$x_c = 0.126$ ($\rho_c = 1.3$ g/cm$^3$), $T_c = 12550$ K, $P_c = 1.3$ GPa.
\begin{figure}
\centering 
\includegraphics[width=0.6\columnwidth]{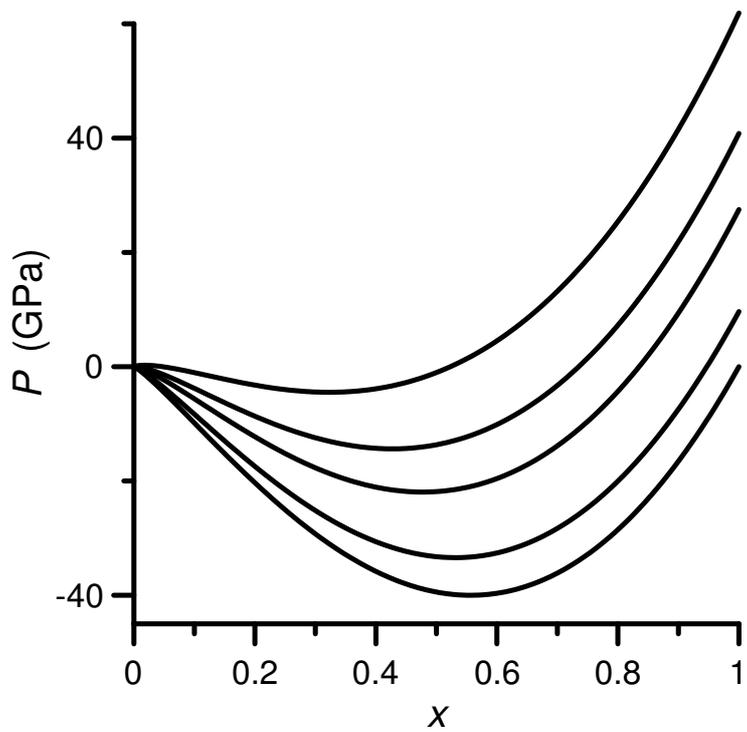}
\caption{Isotherms of molybdenum at low temperatures. Temperatures down up are 0, 2000,
5000, 7000, 10000 K.}\label{Fig:Tlow}
\end{figure}
In figure \ref{Fig:Tlow} some isotherms near the zero isotherm are presented. Internal energy and pressure
presented were used in the hydrodynamical code to investigate the ablation motion of a thin
(10 nm thickness) molybdenum film on the glass substrate (glass thickness is 500 $\mu$m).

\section{Interference phenomena in the ablation of molybdenum film on the glass}
When the thin molybdenum film placed on the glass undergoes an ultrashort laser pulse, it breaks
down at sufficient irradiation fluence. Detached part of the film moves away from that remaining
on the substrate with the velocity $u_f$ while heated and pressurized film on a glass generates the
shock wave propagating into the glass with the velocity $u_s$. In addition the remainder of the
film, initially lying on the glass, begins to separate away from the substrate. This multilayer
distribution of a matter creates the interference pattern in the form of Newton rings when the
probe laser pulse is exposed to both the metal film side as well as to the glass side. We consider
the interference phenomena in the framework of Fresnel theory. Acording to the experimental
conditions we consider only the light incidence in the direction, normal to the metal surface. So
the difference between S- and P-polarization of light disappears. For definiteness we consider
P-polarization, when the magnetic field vector is perpendicular to the plane of the light of
incidence. Vacuum wave length is equal to $\lambda=527$ nm. Boundary conditions for the electric
and magnetic field strength then connect the magnetic field strength in two optical media. We
designate in $j$-layer $H_{\rm t}(j)$ the magnetic field in the wave propagating in the direction of the
incident wave andd $H_{\rm r}(j)$ in the reflected wave, propagating in the opposite direction. When
optical media $j$ and $j+1$border with each other, then the magnetic field in the medium $j+1$
is expressed through the magnetic field in the medium $j$ as
\begin{eqnarray}
H_{\rm t}(j+1)&=&b_{\rm {tt}}(j-1,j)H_{\rm t}(j)+b_{\rm {tr}}(j-1,j)H_{\rm r}(j)\label{Ht}\\
H_{\rm r}(j+1)&=&b_{\rm {rt}}(j-1,j)H_{\rm t}(j)+b_{\rm {rr}}(j-1,j)H_{\rm r}(j),\label{Hr}
\end{eqnarray}
where
\begin{eqnarray}
\label{btt}
b_{\rm {tt}}(j-1,j)&=&\frac{1}{2}\left(1+\frac{\epsilon_j}{\epsilon_{j+1}}\frac{k_j}{k_{j+1}}\right)\exp[\rmi(k_{j+1}-k_{j})L_{j,j+1}]\\
\label{btr}
b_{\rm {tr}}(j-1,j)&=&\frac{1}{2}\left(1-\frac{\epsilon_j}{\epsilon_{j+1}}\frac{k_j}{k_{j+1}}\right)\exp[-\rmi(k_{j+1}+k_{j})L_{j,j+1}]\\
\label{brt}
b_{\rm {rt}}(j-1,j)&=&\frac{1}{2}\left(1-\frac{\epsilon_j}{\epsilon_{j+1}}\frac{k_j}{k_{j+1}}\right)\exp[\rmi(k_{j+1}+k_{j})L_{j,j+1}]\\
\label{brr}
b_{\rm {rr}}(j-1,j)&=&\frac{1}{2}\left(1+\frac{\epsilon_j}{\epsilon_{j+1}}\frac{k_j}{k_{j+1}}\right)\exp[-\rmi(k_{j+1}-k_{j})L_{j,j+1}]
\end{eqnarray}
In expressions (\ref{btt},\ref{btr},\ref{brt},\ref{brr}) $\epsilon(j)=\epsilon_1 (j)+\rmi\epsilon_2 (j)$ is a complex dielectric permittivity of the medium $j$, and $k_j = k_{1j}+\rmi k_{2j}$ is a complex wave number.
\begin{eqnarray*}
k_{1j}&=&k_0\sqrt{\frac{\sqrt{\epsilon_1^2 (j)+\epsilon_2^2 (j)}+\epsilon_1 (j)}{2}},\\
k_{2j}&=&k_0\sqrt{\frac{\sqrt{\epsilon_1^2 (j)+\epsilon_2^2 (j)}-\epsilon_1 (j)}{2}}.
\end{eqnarray*}
Also $k_0=2\pi/\lambda$ is the vacuum wave number, and $L_{j,j+1}$ is the position of the center of a boundary between $j$ and $j+1$ layers. Begining from the last of N layers, where $H_{\rm t}(j_{\rm N})=1$, $H_{\rm r}(j_{\rm N})=0$, we can iterate equations (\ref{Ht},\ref{Hr}) to obtain the incident and reflected waves in the medium 1.

Dynamics of the molybdenum film on a glass being ablated under the action of femtosecond
laser pulse provides us seven layers over a time interval of about 300 ps. According to the scheme
of experiment the probe pulse is devided onto two components. One of them hits a target from
its metal part, and the second one acts on the glass side. When considering the reflection of
the probe pulse acting onto the metal film, the layers 1–8 are vacuum, detached part of film,
vaporized cavity, crater film, cavity between crater film and glass, glass after shock, glass before
shock, vacuum. In this case we are interesting in the temporal evolution of reflectivity
$$
R(t)=\left|\frac{H_{\rm r}(1)}{H_{\rm t}(1)}\right|^2.
$$
For the second component of a probe pulse hitting a glass we shall consider its transmission
through the arising multilayer structure 1–8 with layers in this case being vacuum, glass before
shock, glass after shock, cavity between glass and crater film, crater film, vaporized cavity,
detached part of film, vacuum. Transmission in dependence on time is
$$
D(t)=\left|\frac{1}{H_{\rm t}(1)}\right|^2 \sqrt{\frac{\epsilon(1)}{\epsilon(7)}}.
$$

\begin{figure}
\centering 
\includegraphics[width=0.57\columnwidth]{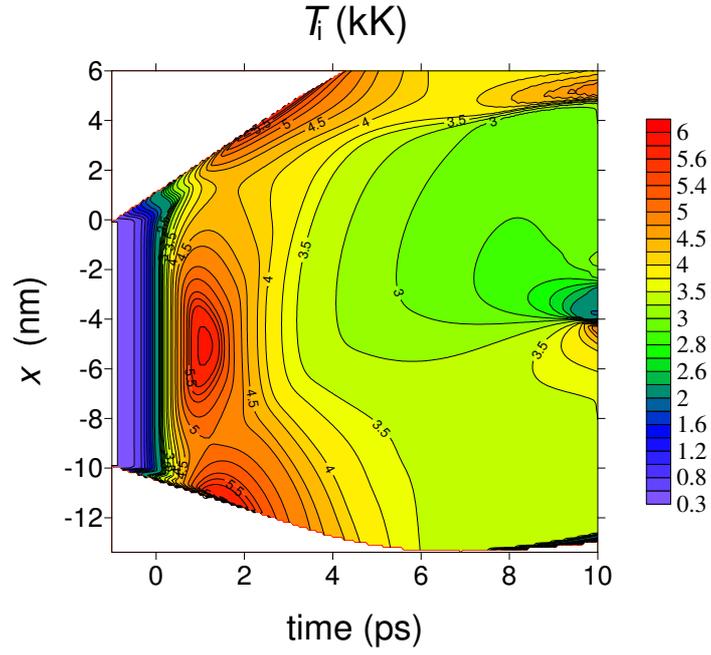}
\caption{The temporal evolution of temperature distribution in the molybdenum film in the
process of its ablation.}\label{Fig:T_xt}
\end{figure}

Application of Lagrangian hydrodynamic code together with thermodynamic and kinetic
characteristics of the film material makes it possible to obtain the parameters of the state at
various points of the film during its motion. Temporal evolution of the ion temperature and its
distribution in thin molybdenum film after the action of femtosecond laser pulse of $t_0 = 402.4$~fs
duration with the value of the absorbed fluence 26.4 mJ/cm$^2$ is presented in figure \ref{Fig:T_xt}. 
Duration $t_0$ corresponds to the gaussian in time profile of impulse intensity, proportional to $\exp(-(t/t_0)^2)$.

\begin{figure}
\centering 
\includegraphics[width=0.54\columnwidth]{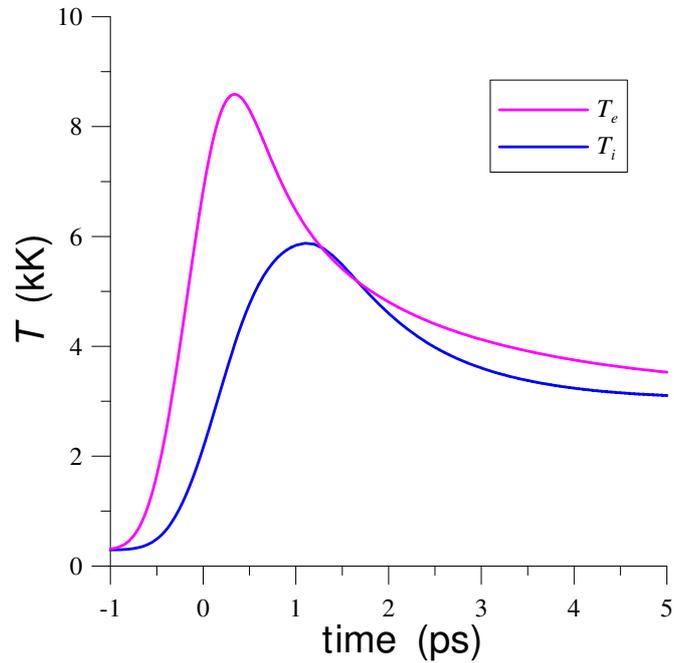}
\caption{Electron and ion temperatures in the center of molybdenum film at an early stage
of its ablation.}\label{Fig:TiTe-x5nm}
\end{figure}

In the time interval shown, the electron and ion temperatures are practically equalized.
Because of thin film under consideration we can assume that the heating of the film is uniform
in its thickness. The process of equalizing the electron and ion temperatures in the central part
of the film is shown in figure \ref{Fig:TiTe-x5nm}.

Cooling of the ionic system is associated with the expansion of the film. Since the ion pressure
is greater, the cooling of the ion subsystem during expansion is more noticeable, and the cooling
of the electronic subsystem is mainly determined by heat exchange with the ionic subsystem.

\begin{figure}
\centering 
\includegraphics[width=0.57\columnwidth]{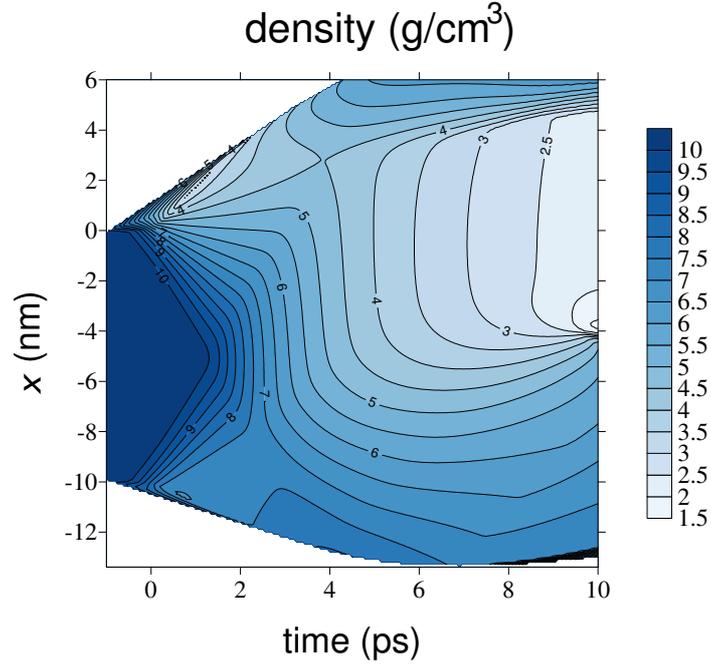}
\caption{The temporal evolution of temperature distribution in the molybdenum film in the
process of its ablation.}\label{Fig:rho_tx}
\end{figure}

The density profile and its temporal evolution with a clearly visible detached part of the
target is shown in figure \ref{Fig:rho_tx}.

Free surface of the film is initially at $x = 0$ and the boundary between the film and glass is
at $x = -10$ nm. Because of the difference in these two boundaries of the film, the asymmetry
of their behavior with time arises leading to the relatively small thickness of the detached
layer of film. We have taken the next parameters of the target material. According to our
hydrodynamical calculations the detached part of the film is about 2.4 nm and moves at the
velocity $u_{\rm f} = 1.0$ km/s at the pump laser irradiation fluence 26.4 mJ/cm$^2$ under consideration.
To calculate evolution of reflectivity and arising pattern of Newton rings in connection with
experiments complex refractive index of this part of molybdennum film was taken $2.4 + 0.1\rmi$,
while for the vaporized cavity between the spalled part and the rest of the film is simply equal to
1. When the rest part of the film adjacent to the glass, in turn tears of from the glass, it moves
at a speed of 0.14 km/s. It can be represented as consisting of two different parts: one, adjacent
to the vapour cavity, having the thickness 2 nm with the same refractive index $2.4+0.1\rmi$ and the
other part, of the thickness 9.1 nm, having the refractive index of about $5.1 + 3.7\rmi$. The shock
wave propagates into the glass with the speed $u_s = 5.8$ km/s. The cavity between the upper
subfilm and the bottom subfilm has refractive index equal 1 as well as the cavity between the
bottom subfilm and the glass.

\begin{figure}
\centering 
\includegraphics[width=0.57\columnwidth]{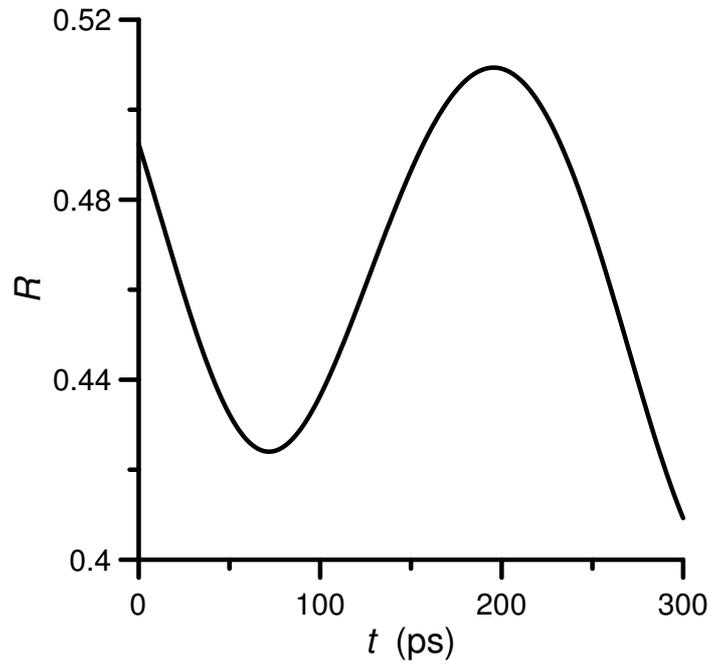}
\caption{The temporal evolution of reflectivity at the ablation of thin molybdenum film
exposed on the glass substrate. Probe laser pulse with the light wave 527 nm acts on the metal
side of a target.}\label{Fig:R_t}
\end{figure}
\begin{figure}
\centering 
\includegraphics[width=0.57\columnwidth]{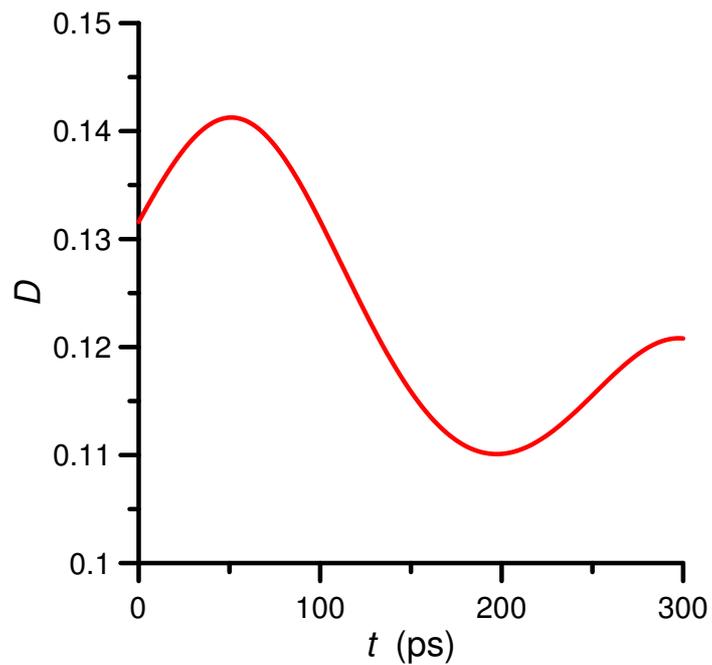}
\caption{The dependence of transmission on time in the process of the ablation of molybdenum
film on the glass. Probe laser pulse acts onto a target from its glass side.}\label{Fig:D_t}
\end{figure}

Temporal evolution of the reflectivity and transmission at the center of the laser spot are
shown in figures \ref{Fig:R_t}  and \ref{Fig:D_t}. Together with the experimentally measured change of reflectivity and
transmission in time this makes it possible to monitor the optical constants of the film matter
during the ablation process.

\begin{figure}
\centering 
\includegraphics[width=0.57\columnwidth]{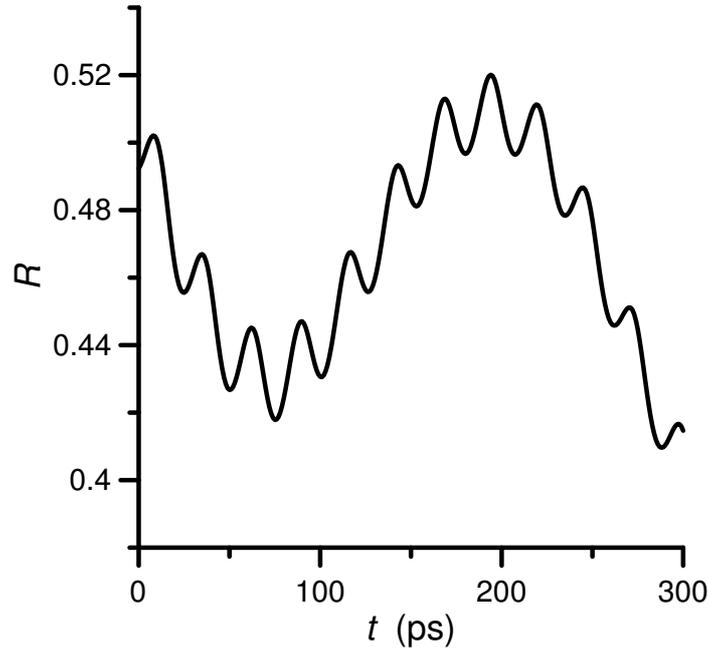}
\caption{The evolution of reflectivity of probe laser pulse in time from the thin molybdenum
film exposed on the glass substrate after the action of pump pulse. Shock wave propagating into
the glass is taken into account. Probe laser pulse acts onto the metal side of a target.}\label{Fig:Rshock}
\end{figure}

\begin{figure}
\centering 
\includegraphics[width=0.57\columnwidth]{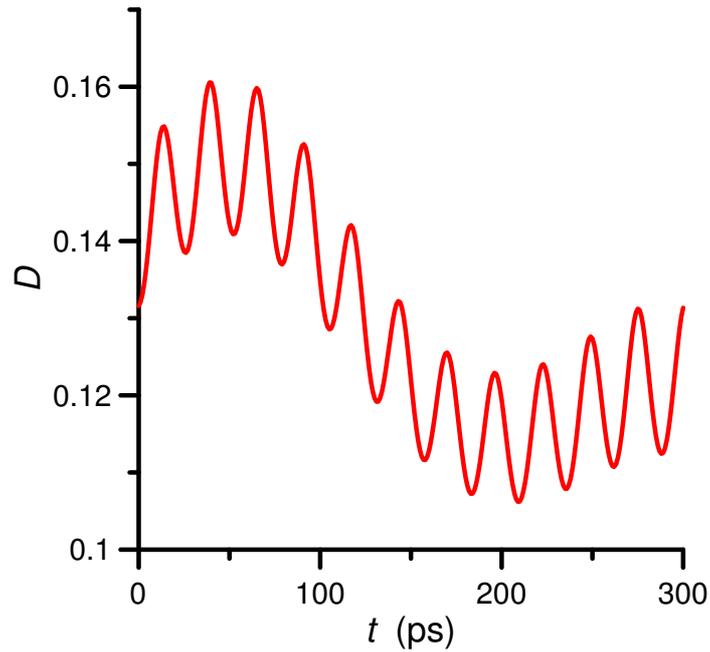}
\caption{Transmission in dependence on time during the ablation of molybdenum film on
the glass. Shock wave in a glass makes a contribution to the transmission. Probe laser pulse
acts onto a glass side of a target.}\label{Fig:Dshock}
\end{figure}

Reflectivity and transmission when taking into account the shock wave in a glass is presented
in figures \ref{Fig:Rshock} and \ref{Fig:Dshock}.
The essential difference between figures \ref{Fig:R_t} and \ref{Fig:Rshock} as well as between \ref{Fig:D_t} and \ref{Fig:Dshock}
shows the possibility of the shockwave monitoring during the pump–probe experiment.

\section{Comparison with experiment}
\begin{figure}
\centering 
\includegraphics[width=0.57\columnwidth]{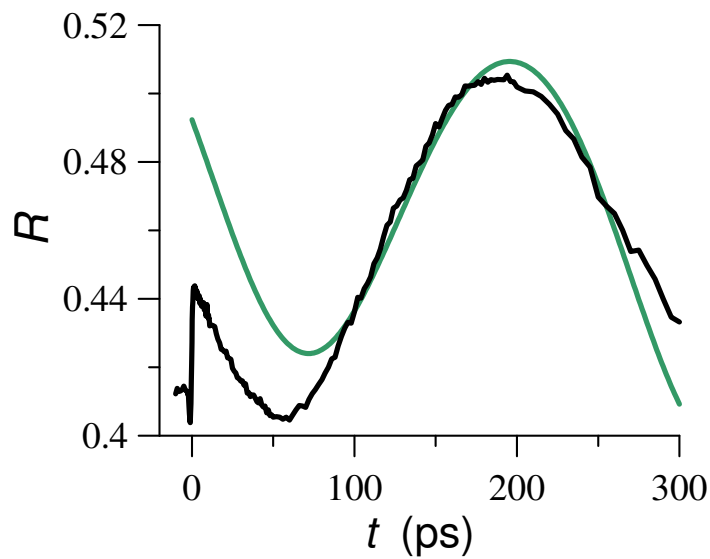}
\caption{Comparison of calculations (green curve) and experiment (black curve), see text
for explanations. The pulse is short, its duration is 0.4 ps. Its maximum is placed at the instant
$t = 0$ at the time axis. The short parts before the instant $t = 0$ relates to the unperturbed
situation. In this situation a reflective index $R$ shown at the vertical axis is $\approx 0.4$. Pulse is
short, as was said, it sharply changes optical parameters of a film. The jump near the point
$t = 0$ corresponds to transition to 2T state. 2T stage lasts few picoseconds, see figure \ref{Fig:TiTe-x5nm}.
Additional efforts are necessary to adjust calculated 2T jump to the observation. We have to
mention that the visibility of interference picture is rather low.}\label{Fig:R}
\end{figure}

Experimental measurements of evolution of the Newton rings were made using technique
described in paper \cite{Domke:2015:JLMN}. Comparison of these measurements with simulated evolution is shown
in figure \ref{Fig:R}.
In experiment initial thickness of a Mo film was 10 nm, incident fluence was
$F_{\rm inc} = 77$ mJ/cm$^2$, duration of pulse was 0.4 ps.

In calculation parameters were: initial thickness of a Mo film was 10 nm, absorbed energy was
$F_{\rm abs} = 26.4$ mJ/cm$^2$, reflection coefficient for 10 nm Mo film placed above thick glass substrate is
0.413, reflection coefficient from a thick Mo film is 0.577. A film absorbs part of incident energy.
Another part reflects back into vacuum and there is also a part which transits into glass. Pulse
duration was the same as in experiment.

Simulation of the film ablation (figures \ref{Fig:T_xt} and \ref{Fig:rho_tx}) show that a film separates into two subfilms.
The upper one in figure \ref{Fig:rho_tx} is thinner than the subfilm contacting with the glass substrate
below. Velocity of the upper subfilm approximately corresponds to experimental velocities. Both
subfilms have reduced densities in comparison with the initial density of the film and are
significantly heated. Therefore their optical parameters differ from those in normal conditions.
Figure \ref{Fig:R} shows that the using of optical constants given above, for the resulting multilayer
structure, makes it possible to obtain satisfactory agreement between the experimental and
theoretical results.

\section{Conclusion}
We have considered the behaviour of the thin molybdenum film on a glass substrate after the
action of femtosecond laser pulse. The subjects of our study are two-temperature thermodynamic
functions of molybdenum, dynamics of the ablation of the film and the optical response to the
action of probe laser radiation. Presented model makes it possible to investigate all aspects of
this problem.

\ack
The work concerning thermodynamics and kinetical properties of molybdenum is supported by
grant No. 16-02-00864 from the Russian Foundation for Basic Research. The work on numerical
simulation of heating and expansion of ultrathin molybdenum film is supported by Federal
Agency for Scientific Organizations (FASO Russia), project No. 0033-2018-0004.
he authors (JW, CA, SR, HPH) would like to thank the company Plansee SE in Austria for providing the Mo thin film samples. The authors also gratefully acknowledge the financial support by the Deutsche Forschungsgemeinschaft (DFG) (HU 1893/2-1). This work was also partly funded by the Erlangen Graduate School in Advanced Optical Technologies (SAOT) by the DFG in the framework of the German excellence initiative.

\section*{References}
\bibliographystyle{unsrt}
\bibliography{lit_Mo}

\end{document}